%% file: ms.tex
\documentclass[12pt,preprint]{aastex}

\begin{document}

\title{ The Solar Heavy Element Abundances: I. Constraints from Stellar Interiors.}

\author{Franck Delahaye $^{1,2}$ and M.H. Pinsonneault $^1$ \\
 1 Department of Astronomy, The Ohio State University, Columbus OH 43210 USA \\
 2 LUTH, (UMR 8102 associ\'ee au CNRS et \`a l'Universit\'e Paris 7),
   Observatoire de Paris, F-92195 Meudon, France.}

\begin{abstract}
 
The latest solar atmosphere models include non-LTE corrections and 3D hydrodynamic convection simulations.  These models predict a significant reduction in the solar metal abundance, which in turn leads to a serious conflict between helioseismic data and the predictions of solar interiors models. We demonstrate that the helioseismic constraints on the surface convection zone depth and helium abundance combined with stellar interiors models can be used to define the goodness of fit rigorous for a given chemical composition.  After a detailed examination of the errors in the theoretical models we conclude that models constructed with the older and higher solar abundances are consistent (within $2 \sigma$) with the seismic data. However, models constructed with the proposed new low abundance scale are strongly disfavored, disagreeing at the $15 \sigma$ level.  We then use the sensitivity of the seismic properties to abundance changes to invert the problem and infer a seismic solar heavy element abundance mix with two components: meteoritic abundances, and the light metals CNONe.  Seismic degeneracies between the best solutions for the elements arise for changes in the relative CNONe abundances and their effects are quantified.  We obtain $Fe/H = 7.50+/-0.045+/-0.003(CNNe)$ and 
$O/H = 8.86+/-0.041 +/-0.025(CNNe)$ on the logarithmic scale where H = 12 for the relative CNNe mixtures in the GS98 mixture; the second error term reflects the uncertainty in the overall abundance scale from errors in the C,N, and Ne abundances relative to oxygen. These are consistent within the errors with the previous standard solar mixture.  However, the inferred solar oxygen abundance is in strong conflict with the low oxygen abundance inferred from the 3D hydro models.  Changes in the Ne abundance can mimic changes in oxygen for the purposes of scalar constraints.  However, models constructed with low oxygen and high neon are inconsistent with the solar sound speed profile.  The implications for the solar abundance scale are discussed.  

\end{abstract}


\section{Introduction}

Despite its familiarity, the Sun retains the capacity to surprise astronomers and challenge our understanding of stellar physics.  The excellent agreement between helioseismic data and theoretical predictions has been, until recently, both one of the most stringent tests of stellar interiors models and one of the greatest successes of stellar theory.  Theoretical predictions of the internal structure of the Sun, however, are sensitive to the internal opacity and thus the solar metal abundance.  The concordance between helioseismology and stellar interiors models is therefore restricted to a relatively narrow range in solar heavy element abundances. The standard solar mixture (Grevese \& Sauval 1998, hereafter GS98) lies comfortably within the permitted range for concordance.

Recent proposed reductions in the bulk metallicity of the Sun from a new generation of theoretical stellar atmospheres, however, drastically degrade the agreement between helioseismic data and theoretical predictions of the internal solar structure.  The claimed reduction in the solar metallicity is modest for the heavier metals (~0.05 dex) and substantial (0.13 to 0.23 dex) for the lighter metals (CNONe).  In order to understand the origin of these changes, a brief review of solar abundance measurements is warranted.  The relative abundances of the heavier metals (such as iron and silicon) in the protosolar nebula can be inferred with high precision in meteorites.  Because the solar hydrogen abundance is not measured in meteorites, inferring the absolute metal abundances requires accurate measurements of photospheric abundances and the usage of stellar atmospheres theory.  To complicate matters further, even the relative abundances of the lighter metals (e.g. CNO) cannot be inferred from meteorites, and solar abundances of these elements rely solely on photospheric measurements.  The abundances of noble gasses cannot be directly measured in either meteorites or the photosphere; they must be inferred from solar wind measurements which are subject to complex systematic effects.  Although helioseismology can be used to infer the surface helium abundance, the neon abundance is both potentially important and difficult to measure precisely.

Standard stellar atmospheres theory makes some important simplifying assumptions in inferring abundances from the measured equivalent widths of spectral lines.  The thermal structure is usually derived assuming a thin atmosphere in hydrostatic equilibrium, which is an excellent approximation for the Sun.  However, there are two other major assumptions that may be more problematic.  Horizontal temperature variations (from granulation) are neglected in standard stellar atmosphere calculations, and the level populations as a function of optical depth in the atmosphere are assumed to be in LTE (i.e. they can be derived from the local temperature, density, and abundance alone).  The presence of nonzero velocity fields in the atmosphere is calibrated out with an ad-hoc microturbulence parameter that is tuned to yield abundance estimates independent of equivalent width for a given ionization and excitation state.

The proposed reduction in the solar metallicity is tied directly to relaxing the above assumptions in the model atmospheres.  Non-LTE effects will tend to boost the level populations for high excitation states, and therefore lower the inferred abundances derived from a given equivalent width.  Numerical 3D simulations of stellar convection (Asplund et al. 2005, AGS05) can be used to study the effects of convection on both the thermal structure and the magnitude of horizontal temperature fluctuations.  These simulations have claimed that horizontal temperature fluctuations are substantial across a wide range of optical depths well above the top of the convection zone inferred from mixing length theory.  The mean thermal structure in these simulations differs dramatically from traditional stellar atmospheres calculations, in the sense that the outer layers of the atmosphere are significantly cooler.  Both of these changes have the net effect, for most species, of boosting the predicted average populations of the species and excitation levels used to measure abundances (a direct effect) and of lowering the continuum opacity (an indirect effect).  The overall result is once again a reduction in the abundances inferred from a given equivalent width.  Significantly, the new models reproduce the line shapes and widths of lines formed relatively deep in the photosphere without requiring an ad hoc microturbulence parameter.

The possible revision of the solar abundance scale has triggered a burst of activity from solar modellers (see, for example, Bahcall \& Pinsonneault 2004; Bahcall, Serenelli, \& Pinsonneault 2004; Bahcall, Serenelli, \& Basu 2005a,b; Basu \& Antia 2004a,b; Antia \& Basu 2005; Turck-Chi\`eze et al. 2004; Guzik \& Watson 2004; Guzik, Watson, \& Cox 2005; Seaton \& Badnell 2004; Badnell et al. 2005; Montalban et al. 2004).  From the papers cited above, there is a consensus that stellar interiors models constructed with the new mixture are incompatible with helioseismic data.  Different solutions have been proposed, some implying changes in the input physics like opacities, diffusions and some proposing extreme changes in the abundance of Neon (Antia \& Basu  2005; Bahcall, Basu \& Serenelli 2005).  A common theme has been the implied assumption that the revised solar abundances are correct, and that the problem must lie in the physics of the stellar interiors models.

In this paper we propose another approach. In our view this problem is not a conflict between theory and observation.  Instead, the theory of stellar atmospheres and the theory of stellar interiors cannot be simultaneously reconciled with helioseismic data.  In this paper we critically analyze the best solar abundances inferred from a combination of helioseismic data and stellar interiors theory.  In Paper II (Pinsonneault \& Delahaye 2005, in prep) we discuss the internal consistency and uncertainties in solar abundances inferred from stellar atmospheres theory.  We begin in section 2 with a discussion of the helioseismic constraints on solar models, with a focus on the observational measurements of the surface helium abundance $Y_{surf}$ and convection zone depth $R_{CZ}$.  In section 3 we turn to the stellar interiors models and their uncertainties.  We first derive the uncertainties and central values of $Y_{surf}$ and $R_{CZ}$ as a function of the solar abundances, and then map these into inferred best values for the heavy element abundances in the Sun.  This exercise quantifies the problem with a reduced solar metallicity from an interiors perspective, as well as distinguishing which of the abundances are actually constrained with the seismic data. In section 4 we discuss our results.

\section {The Helioseismic Constraints}

The goal of this paper is to evaluate the uncertainties in the solar heavy element abundances from both stellar interiors and atmospheres theory.  The most important physical effect that metals have on solar structure is their contribution to the opacity in the radiative interior.  The CNO elements are abundant and contribute substantially to the Rosseland mean opacity at temperatures of a few million K, primarily from bound-free transitions.  As a result, their abundance can affect the depth of the solar surface convection zone and the thermal structure of the outer layers of the solar radiative core.  However, they are fully ionized at higher temperatures and have little impact on the central temperature or thermal structure of solar models.  The heavier metals (Mg, Si, Fe) are less abundant and make a smaller contribution to the opacity at temperatures of a few million K than the CNO species do.  However, they retain bound electrons to much higher temperatures and some of them (especially iron) are important opacity sources even at the center of the Sun. Neon is intermediate in behavior between the two classes of behavior described.

From the brief discussion above, changes in the abundance of different elements will impact different regions in the solar interior.  Helioseismology is uniquely capable of distinguishing between the effects of changing the abundances of light and heavy metals because it provides diagnostics of the thermal structure of the bulk of the interior of the Sun.  The most commonly used diagnostic is the sound speed as a function of depth, but there are also precise scalar constraints on the depth of the solar surface convection zone and the surface helium abundance.  In our view these scalar constraints capture enough of the information encoded in the seismology to permit a rigorous estimate of the metallicity that is consistent with stellar interiors physics.  We therefore begin with the sound speed profile before analyzing the best current seismic constraints on the convection zone depth and helium abundance, along with their observational errors.

\subsection{The Speed of Sound as a Function of Depth}

The Sun is an acoustic cavity with a rich spectrum of oscillation frequencies; more than $10^5$ distinct modes have been identified.  Different modes penetrate to different depths, and information from these frequencies can be inverted to obtain an estimate of the sound speed as a function of depth.  A nice introduction to the theory of oscillations can be found in Christensen-Dalsgaard \& Berthomieu (1991).  In the same volume the inversion problem is treated by Gough \& Thompson (1991).  The usual procedure is to use a reference solar model to predict a set of frequencies.  Differences between the observed and predicted frequencies can be used to solve for differences in structure between the actual Sun and the reference model.  The results are insensitive to the choice of reference model (Basu, Pinsonneault, \& Bahcall 2000).  A combination of a limited number of modes and a short crossing time in the core implies that the inverted sound speed profiles are only reliable outside 0.05 solar radii, and the interpretation of the outer layers is complex because a variety of physical effects must be considered.  However, the seismic data can be used to provide a stringent test of solar structure for the bulk of the radial extent of the Sun.  

In Figure 1 we compare the inverted solar structure from Basu, Pinsonneault, \& Bahcall (2000) with the theoretically predicted sound speed profiles for solar models constructed with the GS98 and AGS05 composition mixes; the input physics and assumed solar properties can be found in Section 3.1.  Differences are defined in the sense (Model-Sun)/Sun.  It is immediately apparent that the GS98 model is close to the real Sun and that the AGS05 model is discrepant.  The statistical significance of these deviations, however, is less easy to determine.  It is appealing in principle to use the goodness of fit in diagrams like this as a quantitative measure of agreement.  However, it is important to remember that the deviations from the real Sun in a figure such as this are strongly correlated. For example, in a solar model of fixed radius and mass a density excess at one point necessarily implies a density deficit elsewhere.  In addition, a proper weighting of the errors would have to account for the theoretical errors as a function of depth, which is a nontrivial exercise.  We therefore will use diagrams such as Figure 1 to illustrate the impact of changes in the input physics, but will concentrate on reproducing scalar constraints for the purposes of determining the solar abundance mix most consistent with seismic data.  The most significant exception is the solar neon abundance, which we will return to in Section 3.2.  Changes in the neon abundance affect the goodness of fit between the core and surface, and the sound speed profile can thus be used to place additional constraints beyond those from scalar constraints alone.

\subsection{The Depth of the Surface Convection Zone}

Theoretical solar models have radiative cores and convective envelopes.  This prediction is confirmed by helioseismic data, which is also able to pinpoint the transition between the two to high precision.  The basic diagnostic employed in modern studies is the gradient in the sound speed, following the general approach of Gough (1986).  A discontinuity in the temperature (and, by extension, the sound speed) gradient between an effectively adiabatic deep convective envelope and a radiative core is both predicted and observed (see Christensen-Dalsgaard, Gough, \& Thompson 1991). The most current estimates for the fractional depth of the transition point, in solar radii, is 0.7133 +/- 0.0005 (Basu \& Antia 2004), which is effectively identical to the 1991 estimate of 0.713 +/- 0.003.  For a given reference model the random error in the convection zone depth is small, of order $5 \times 10^{-4}$.  Changes in the envelope heavy element abundances from GS98 to AGS05 produce no change in the seismic estimate (Basu \& Antia 2004).

Systematic effects are still small, but larger than the random errors.  Basu (1998) found that reference models which are poor fits to the seismic data can change the inferred convection zone depths by +/- 0.0015.  Because these reference models are of lower quality (on other grounds) than the ones used to derive the central value of $R_{CZ}$, we choose to treat this as a three sigma effective systematic error.  Our adopted value for $R_{CZ}=0.7133 \pm 0.0005 (rand) \pm 0.001(sys)$, for a total error of 0.0011.

\subsection{The Surface Helium Abundance}

It is well known that the adiabatic gradient is decreased in ionization regions.  Gough (1984) recognized that this could be used to infer the helium abundance in the solar convection zone.  A number of investigators have subsequently obtained estimates of $Y_{surf}$ using different methods and choices of the equation of state.  This is a more difficult problem than the convection zone depth because it is more sensitive to the choice of equation of state.  Some (but not all!) authors obtain systematically lower abundances for the MHD equation of state than for the OPAL equation of state.  In addition, other effects which become unimportant at deeper layers, such as non-adiabatic corrections, may need to be accounted for at the very shallow depth of the helium ionization zone.  An increase in the inferred $Y_{surf}$ (from 0.244 to 0.248) was reported by Basu \& Antia (2004) when the lower AGS05 heavy element abundances were used.  This indicates that the envelope metal abundance may have some impact on the observed $Y_{surf}$, in the sense that it would worsen agreement between the AGS05 model (which favors lower helium abundances).  In the interests of placing conservative error estimates (and the absence of other studies confirming the effect) we neglect this potential effect and treat $Y_{surf}$ as independent of the assumed metal abundances.

The average between MHD and OPAL results for Baturin \& Ayukov (1997), Kosovichev (1997), Basu (1998), Richard et al. (1998), Di Mauro et al. (2002) and Brun et al. (2002) are respectively 0.24, 0.24, 0.25, 0.245, 0.25, and 0.252.  The first two had substantial differences (opposite in sign) between MHD and OPAL solutions and were not published in the refereed literature; the last represents a different method of obtaining surface helium (consistency in the model and inferred density profiles).  As a result, we use Basu, Richard et al., and Di Mauro et al. (with estimates from both equations of state) to infer a mean and dispersion.  The result is a mean abundance of 0.2483 with a standard deviation of 0.0043.  If we had used all sources the mean value would be 0.2462 and the standard deviation would be 0.0084; the mean difference between MHD and OPAL is in the three adopted references is -0.006, indicating the errors are primarily systematic in nature.  We therefore adopt $Y_{surf} = 0.2483 \pm 0.0046$.

\section{Solar Abundances from Helioseismology and Interiors Models}
 
We wish to infer the solar heavy and light metal abundances required for theoretical models that reproduce the observed solar surface helium and convection zone depth.  There are three essential steps involved in determining a seismic solar metallicity and its associated error.  First, the observed solar properties and their errors must be obtained.  Second, errors in the input solar model physics introduce uncertainties in the theoretically predicted surface helium and convection zone depth.  These uncertainties can be correlated; for example, increasing the degree of gravitational settling both deepens the model convection zone depth and decreases the surface helium abundance.  At this point we can both define a difference between theory and observation and an associated error in that difference for a given solar composition.  Finally, we can determine the impact of changes in the solar abundances upon the seismic properties of the models.  This would appear to be an underdetermined problem, since there are 17 heavy elements included in the OP opacity calculations, but only two constraints.  However, we will demonstrate that the abundance problem can be treated as one with three principal components (the meteoritic abundances, oxygen, and the neon to oxygen ratio).  Changes in the abundance of the heavier metals primarily affect the surface helium abundance, while changes in the CNO abundances primarily affect the surface convection zone depth.  As a result, we can solve for the heavy (meteoritic) and light (photospheric) abundances consistent with the solar data as a function of the assumed Ne/O.  We therefore begin this section by determining the theoretical errors in $R_{CZ}$ and $Y_{surf}$, and follow with a determination of the solar abundances consistent with seismic data.

\subsection{Errors in Theoretical Interiors Models}

In order to determine the uncertainties in $R_{CZ}$ and $Y_{surf}$, we began by constructing a reference model that is used to obtain the central values for $R_{CZ}$ and $Y_{surf}$.  We then constructed a series of other solar models in which one parameter at a time is modified. Some error sources are random in nature, and we denote these accordingly along with the associated one $\sigma$ errors.  In other cases the underlying errors are systematic; examples include the choice of equation of state and the quantum mechanical calculations of the Rosseland mean opacity at a given density, temperature, and composition.  In this case we adopt the ``effective two sigma'' approach from previous work, treating the difference between independent calculations as being equivalent to two sigma random errors.  Unlike prior work, we incorporate information about correlated changes in the two seismic variables in our error estimate.  The net result is a robust theoretical estimate of the errors in seismic properties for a given solar composition.  Our reference model is described in section 3.1.1, the construction of the OP opacity tables is described in section 3.1.2, and the theoretical error budget is defined in section 3.1.3.

\subsubsection{Reference solar model}

We used the Yale Rotating Evolution Code to generate solar models from the zero age main sequence to the solar age. The mixing length and initial helium abundance are adjusted to reproduce the observed solar luminosity ($L_{\odot}$), radius ($R_{\odot}$) and surface $Z/X$ ratio at the present epoch. For the purposes of inferring the theoretical errors, we adopted the Grevesse \& Sauval (1998) mixture and we varied individual ingredients to establish the sensitivity of the results to the assumed solar properties and input physics.  Our base case has $L_{\odot} = 3.8418 \times 10^{33}~erg~s^{-1}$, $(R_{\odot})= 6.9598 \times 10^{10}~cm$, and an age of 4.57 Gyr (taken from a ZAMS starting model.  We use the OPAL 2001 equation of state (Rogers, \& Nayfonov 2002).  For our opacities we use OP data above $10^4 K$ (Badnell et al. 2005).  (See the next section for the procedure used to construct the opacity tables, which differs from other studies because of the smaller number of species used in OP than in the previous generation of OPAL tables).  For lower temperatures we use Alexander \& Fergusson (1994) molecular opacities.  For our boundary condition we use the Krishnaswamy $T-\tau$ relation and the standard mixing length theory for convection.  Nuclear reaction rates are the same as Bahcall \& Pinsonneault (2004), adopting the lower ${14}N+p$ cross-section of 1.77 keV from Angulo \& Descouvemont (2001) for the base case (further discussion below).  Errors in the cross-sections are taken from Adelberger et al. (1998), and the central values for the main solar nuclear reactions are close to those in that paper with the exception of a lower adopted pp cross-section of $3.94 \times 10^{-22}$ keV taken from the NACRE compilation (Angulo et al. 1999).  We used weak screening for our nuclear reaction rates.

For gravitational settling we use the Thoul et al. (1994) method, with the coefficients computed for elements heavier than helium as if all elements settled at the same rate (taken to be that for fully ionized iron).  The effective diffusion coefficient was reduced by a scale factor to simulate the effects of mixing. The potential reduction in metal diffusion from radiative levitation and partial ionization was considered separately.  The seismic properties of our best choices of input physics for the two composition mixtures that we consider are ($R_{CZ}$,$Y_{surf}$) of (0.71558,0.24722) for GS98 and (0.72985,0.23304) for AGS05 respectively.  The GS98 mixture will be the basis for the parameter variations that follow, and both it and the AGS05 mixture will be compared to the seismic data after we establish both the construction of the new opacity tables and the error budget.

\subsubsection{Opacity tables}

The new OP opacities provide a welcome test of the theoretical uncertainties in opacity calculations.  Some care, however, is required when comparing OP results to those from the OPAL group. While the OP data include 17 elements (H, He, C, N, O, Ne, Na, Mg, Al, Si, S, Ca, Ar, Cr, Mn, Fe and Ni) OPAL includes 21 species (adding P, Cl, K and Ti to the OP list).  In order to infer the best means of correcting for this difference, we proceeded as follows, using the OPAL data as a test. Our base case (compOPAL) included all 21 elements.  We then constructed 17 element Rosseland mean opacity tables from the OPAL data using the GS98 abundances, zeroed out the abundances of the elements missing in OP, and then redistributed the number fraction of P, Cl, K and Ti to the other species in two ways (comp1 and comp2). We then compared with compOPAL to infer the best method for accounting for the missing elements.  

The simplest approach is simply to increase the abundances of all metals by the ratio of the total number fraction of all 21 species to the number fraction of the 17 included in OP.  Because the bulk of metals in the Sun are CNO elements, this procedure has the net effect of redistributing P, Cl, K and Ti opacity to lower temperatures. We call this mixture comp1.

Another approach is to redistribute the number fraction of the 4 extra elements to their closest included neighbor in the periodic table. For this second composition, we have $f_S=f_S(GS98)+f_P(GS98)$, $f_{Ar}=f_{Ar}(GS98)+f_{Cl}(GS98)$, $f_{Ca}=f_{Ca}(GS98)+f_K(GS98)$ and $f_{Cr}=f_{Cr}(GS98)+f_{Ti}(GS98)$.  We call this mixture comp2.

The base values of $R_{CZ}$ and $Y_{surf}$ for compOPAL, comp1, and comp2 are 0.71767/0.24852, 0.71686/0.24703 and 0.71703/0.24759 respectively.  The compOPAL and comp1 differences are thus of the order of 0.13\%/0.6\% for the seismic properties.  The neutrino fluxes are affected at the 1 to 2\% level. The variation are respectively $\lesssim 0.1\%/0.3\%$ for $R_{cz}/Y_{surf}$ and $\lesssim 0.1\%$ for the neutrino fluxes when we compare comp2 to the compOPAL. This clearly indicates the better representation of the original OPAL mixture by the modified 17 elements mixture comp2. While small, these differences are of the order of the observational uncertainties (0.15\% for $R_{CZ}$). We therefore adopt the second method for constructing opacity tables.  The full set of abundances used for the GS98 and AGS05 mixtures in our work are presented in Table 1.

\subsubsection{Details of the Theoretical Error Budget.}

Our theoretical error budget is presented in Table 3.  In column 1 we describe the error source.  Its central value and the adopted error are given in column 2 if the error source is random in nature.  If the error source is systematic, the comparison case is noted here.  Column 3 indicates whether the error is treated as random or systematic. Columns 4 and 5 report the differences in $R_{CZ}$ and $Y_{surf}$ arising from the model ingredient.  For systematic errors, the difference in seismic results is taken as an effective two $\sigma$ result, and the tabulated value is thus taken as half the net change.  Column 6 gives the source of the result; parameter variations from other solar model calculations were included relative to their reference values.  Errors are symmetric unless otherwise noted.  The two main features of Table 3 are the relatively small inferred errors and their broad distribution; no single ingredient is the most important.

Rows 1-3: Initial Conditions

The uncertainties in luminosity (0.4 \%) and age (0.01 Gyr) are taken from BP95 and treated as random errors.  For radius errors we chose to compare the low radius inferred from solar meridian transits of $6.95508 \times 10^{10} cm$ (Brown \& Christensen-Dalsgaard  1998) with the reference value of $6.9598 \times 10^{10} cm$ and treat the error as systematic.  None of these ingredients contribute substantially to the theoretical uncertainties; this would include models that start in the pre-MS phase of evolution, which would effectively yield a solar nuclear age 30-40 Myr younger than the canonical value.

Row 4: Equation of State
 
The choice of equation of state has a small effect on the stellar interiors value of $R_{CZ}$ and a modest one on $Y_{surf}$.  We took the difference between the OPAL96 (Rogers, Swenson, \& Iglesias 1996) and the OPAL01 equation of state as our estimate of the uncertainty arising from the treatment of the EoS.  Comparisons involving the OPAL and MHD equations of state yield even smaller relative differences (Boothroyd \& Sackmann 2003).

Rows 5-7: Nuclear Reaction Rates.
The main energy source for the Sun is the pp chain, and we therefore include the three main pp reactions ($pp=So_{1,1},~He^3+He^3=So_{3,3}$, and $He^3+He^4=So_{3,4}$) in our error budget.  We consider only errors in the cross-section at zero energy (So), as differences in higher order terms have minimal impact for solar models.

For $So_{1,1}$, we used the NACRE cross-section of $3.94 \times 10^{-22} keV$ from Angulo et al. (1999), who did not include a specific error discussion in this theoretically calculated quantity.  For the error budget we used the estimates in Adelberger et al. (1998), adding the systematic and random errors in quadrature to obtain a fractional values of +0.022/-0.013.  

For $So_{3,3} and So_{3,4}$ we adopted the Adelberger et al. (1998) cross-section of $5.40 \pm 0.40 \times 10^3$ and $0.53 \pm 0.05$ keV b respectively.  These are close to the NACRE values of $5.18 \times 10^3$ (no explicit error quoted) and $0.53 \pm 0.09$ respectively.  The net impact of adopting NACRE cross-sections for these reactions would be minimal.

For $So_{1,14}(=N^{14}+p)$, recent changes have been substantial.  For our reference case we adopt $ So_{1,14} = 1.77 keV b$ (Angulo \& Descouvemont 2001), to be compared with the Adelberger et al. value of 3.32.  More recent papers (Runkle et al. 2005; Imbriani et al. 2005) yield concordant measurements of $1.68 \pm 0.09 (stat) \pm 0.16 (sys)$ and $ 1.61 \pm 0.08 (stat)$ respectively; as a result we have shifted our base case to include the lower value.  The revised uncertainties make the CNO cycle an even smaller contributor to the solar luminosity (at the 0.8 \% level) and a negligible portion of the error budget. 

Row 8: Low Temperature Opacities.

For the uncertainties in low temperature opacities we used the differential effect reported by Boothroyd \& Sackmann (2003) when the Sharp (1992) opacities were used in place of the Alexander \& Ferguson (1994) ones and treated the difference as a systematic error. 

Row 9: High Temperature Opacities.

For the high temperature opacities we used the difference between the OPAL96 and OP05 opacities as an estimate of the impact of differences in the opacity at fixed temperature, composition, and density on solar model properties.  We note that the derived uncertainties are substantially smaller than those required to explain the solar convection zone depth problem for the new mixture.

Row 10: Diffusion and Mixing.

We have introduced a multiplicative factor in the settling coefficient to take into account the error in the gravitational settling coefficient and to account for the rotational mixing. The gravitational settling coefficients are calculated using the Thoul's subroutine (Thoul, Bahcall \& Loeb 1994) for which 15\% uncertainties are quoted by the author.  These error estimates are consistent with the good agreement reported relative to the independent calculations of Turcotte et al. (1998).

There is clear evidence, however, that pure diffusion models overestimate the degree of gravitational settling and thermal diffusion in the Sun.  It has long been known that the photospheric lithium abundance is strongly depleted relative to the meteoritic value (Greenstein \& Richardson 1951).  Furthermore, the lithium abundance of young solar analogs on the main sequence is close to the meteoritic value (for example, Soderblom et al. 1993); this implies directly that lithium depletion must occur on the main sequence.
Lithium is easily destroyed in stellar interiors, so the most likely explanation is mild envelope mixing.

The depth of mixing is more controversial, but there is evidence from beryllium that suggests that deep mixing is unlikely.  It was long thought that the less fragile element beryllium was also depleted in the Sun by a factor of around two (Ross \& Aller 1974).  The only accessible beryllium lines, however, are in the crowded ultraviolet portion of the spectrum.  More recently, there have been claims that the continuous UV opacity in the Sun has been underestimated (Balachandran \& Bell 1998; Asplund 2004).  The evidence presented is that the solar oxygen abundance inferred from UV lines is too small without an increase in opacity, and there has been plausible evidence that such an increase could come from numerous weak iron lines.  We will return to this point when discussing the solar oxygen in section 4.  If the continuous opacity is underestimated, the line opacity will also be underestimated; as a result, the corrected beryllium abundance is close to the photospheric value.  This has frequently been misunderstood in the literature  and it has been claimed that beryllium must be undepleted. In fact, the errors in the ad hoc increase in continuous opacity are
large enough to permit modest beryllium depletion (at the 0.2 dex level).  In either case, however, the inferred depletion is small enough to indicate that any extra mixing is mild and relatively shallow.

Meridional circulation and shear instabilities arising from differential rotation have been demonstrated to have sufficient energy to drive the required degree of mixing (see Pinsonneault et al. 1989; Pinsonneault 1997).  Mixing reduces local composition gradients, while settling increases them.  The net effect is a modest reduction in the degree of gravitational settling inferred from the models (Chaboyer, Demarque \& Pinsonneault 1995; Richard et al. 1996; Brun, Turck-Chi\`eze \& Zahn 1999; Bahcall, Pinsonneault, \& Basu 2001).  Different physical models for rotational mixing predict similar degrees of reduction in the efficiency of rotational mixing in models that reproduce the solar lithium depletion.  Bahcall, Pinsonneault, and Basu (2001) reported a 21 \% reduction in the effective diffusion coefficient, while other published estimates range between 15 \% and 25 \%.  We therefore adopt a central value of 0.2 +/- 0.05 for the reduction in efficiency of settling from mixing.  When
combined with the error of 0.15 in the diffusion coefficients themselves, we therefore adopt a gravitational settling coefficient of 0.8 +/- 0.16 relative to the Thoul et al. (1994) prescription.

Row 11: Differential Settling

The degree of gravitational settling of helium is observationally constrained, which limits the uncertainty in metal diffusion to differential effects.  Radiative levitation can mildly decrease metal diffusion in solar models, while relaxing the treatment of all metals as fully ionized iron can increase it.  The net effect is small.  Turcotte et al. 1998 found that the average metal diffusion rates accounting for both processes were typically close to the bulk metal diffusion rate (of order 10\%) reported by Bahcall, Pinsonneault, \& Wasserburg 1995. Individual results for species agree very closely with their model C and differ at up to the 1.6 \% level for their model H.  We therefore adopt a metal diffusion coefficient error relative to helium diffusion at the 10 \% level as a measure of the selective settling error.

Row 12: Numerical Effects

In this column we include the numerical error in the location of the convection zone depth from differences in the treatment of opacity interpolation, using the Garching and Yale codes for comparison. See Bahcall,~Serenelli \& Pinsonneault  (2004) for a discussion.

\subsubsection{Theoretical Errors in the Seismic Properties of Solar Models.}

We are now ready to quantify the total theoretical error in stellar interiors models and compare our two base cases with the observational data.
Treating the individual changes in the two seismic variables as being perfectly correlated, we have defined an error ellipse for $R_{CZ}-Y_{surf}$. We constructed the covariance matrix for each source and summed them all to obtain a total error. The resulting uncertainties in $R_{CZ}$ and $Y_{surf}$, when treated separately, are $\sigma_{R_{CZ}}=0.0027$ and $\sigma_{Y_{surf}}=0.0032$. The resulting error ellipse is plotted in Figure 2. In the $R_{CZ}-Y_{surf}$ plane, we have represented the helioseismic data with the observational error-ellipse, and the theoretical error-ellipse placed at two different points for our two different mixtures. 

As noted by other authors, the agreement between the model using the GS98 mixture and the helioseismic data is good. Note that our central value is different from previous results reported by one of the authors (MP) because we choose the initial model to include some mixing. We can now see that the results using AGS05 differ not only by many $\sigma_{observation}$ but also by many $\sigma_{theory}$; GS98 differs by less than 1 $\sigma_{theory}$ while AGS05 is ruled out at the 6 $\sigma_{theory}$ level. The advantage of this representation is to show directly that any changes in the input physics has to be extreme in order to reconcile the new composition with the helioseismic data. Furthermore, it illustrates the powerful additional constraints imposed by requiring that both the surface convection zone depth and helium abundance be reproduced.  We use the change in opacity data and settling/mixing as examples for the limited effects of input changes to clear the inconsistency generated by the AGS05 mixture; the vector changes induced by these error sources are indicated in Figure 1 along with the sense of changes in the bulk metallicity.  

Errors in the theoretically calculated opacities have been proposed as a potential explanation of the convection zone depth problem (Bahcall et al. 2005, AGS05).  However, the OP and OPAL groups use independent methods for opacity calculations, and the difference between the two can be used to infer the theoretical uncertainties.  The agreement between OP and OPAL is actually very good, providing some confidence that the opacities at fixed mixture are reasonably secure (Badnell et al 2005).  For solar models the difference does not exceed 4\% across the structure and is less than 2.5\% at the base of the convection zone (Delahaye \& Pinsonneault 2005). Models constructed with OP opacities have deeper surface convection zones, but also increased surface helium abundances; the former effect reduces the difference between theory and observation while the latter effect increases it.  This behavior arises because the OP Rosseland mean cross sections for O is larger than those estimated for OPAL, while the Fe group elements contribution to the Rosseland mean opacity is larger in the case of OPAL compared to OP (see Badnell et al. 2005 for a discussion). As we will show in section 3.2, O acts essentially on $R_{CZ}$ and the Fe-peak elements affects preferentially on $Y_{surf}$.

Asplund et al (2005) also suggested that an underestimation of the settling could also produce a reduction of the depth of the convection zone. In effect, the surface abundances could be small while the interior abundances were higher.  Studies of enhanced settling (Guzic, Watson \& Cox 2005, Montalban et al. 2005) confirmed that an extremely large increase in the settling would be required.  However, there is an intrinsic problem when both helium and convection zone depth are considered. Increasing settling will reduce $R_{CZ}$ and $Y_{surf}$, improving the agreement for the convection zone but worsening it for the surface He abundance.  For the same reason, inclusion of mixing is neutral with respect to the agreement between the new abundance scale and the seismic data.

We therefore conclude that the AGS05 mixture is highly incompatible with the combination of helioseismic observations and solar model interiors physics.  However, composition changes (illustrated in Figure 2 by the green arrow) can clearly restore agreement, which is not surprising given that the sole difference between GS98 and AGS05 is the assumed solar mixture. In the next section we quantify this effect to determine both the best solar mixture of heavy elements and its uncertainty.

\subsection{The Seismic Solar Abundance} 

Now that we have established an error budget for the observational and theoretical prediction of the depth of the convection zone and the surface He abundance, we can convert it into a series of constraints on the compositions if the response of the seismic variables to abundance changes is known.  We begin by arguing that there are three principle components to the solar heavy element mix: elements whose relative abundances can be measured precisely in meteorites, elements that can only be measured in the photosphere, and elements that can only be measured in the chromosphere and corona.  The abundance problem can therefore be reduced to these three components, whose relative abundances can be held fixed but whose absolute value can be permitted to vary.  This physical constraint is strong for the meteoritic elements but weaker for the two other classes. Fortunately, the results for photospheric and coronal abundances can be demonstrated to depend primarily on their most abundant species (oxygen and neon respectively), so the relative variations of other elements (C, N, Ar) will not have a major impact on the derived results.  We then solve for the meteoritic and photospheric abundances consistent within the errors with the seismic data for fixed neon in the following section and consider the effects of varying neon on the derived results afterwards.  Although there is degeneracy in the surface seismic variables between neon and oxygen/iron, we argue that the sound speed profile can be used to demonstrate that high neon/low oxygen mixtures are incompatible with the seismic data taken as a whole.

\subsubsection{Sensitivity to Abundance Changes}

Based on the results of the previous section we have generated an error ellipse for the difference between observation and theory ($R_{CZ}-R^{Sun}_{CZ}$ and $Y_{surf}-Y^{Sun}_{surf}$), combining the observational and theoretical errors. The result is shown in Figure 3.  The solid triangle represents the value for GS98, the solid square AGS05 and the cross the results from Helioseismologie.  We also illustrate the effects of changing different elements and mixtures of elements, defined below.
  
Changing the heavy element abundances will affect the opacity of the mixture. As a consequence, $R_{CZ}$ and $Y_{surf}$ will be altered.  However, each atom has a different influence on the total opacity and we can expect a different response depending on the species or ensemble of ions for which the abundances are modified. With 15 heavy elements in the OP mixture and two observables, the problem is clearly under constrained.

Fortunately, the abundance measurements are not independent of one another.
The relative abundances of the heavier metals can be accurately inferred in meteorites; for our purposes in the OP mixture this includes all elements from Na to Ni.  (Although the noble gas Ar is not measured precisely in meteorites, it is a minor opacity source that does not have a distinct seismic signal.  We therefore simply did not modify its abundance).  The absolute abundance scale, however, must be determined by photospheric measurements, and we treat it as a quantity to be solved for.  We will refer to these as 'meteoritic' abundances, and will use the Fe abundance as a proxy.  Differential settling of metals is a small effect in solar models (Turcotte et al. 1998), and therefore the relative abundances in meteorites should faithfully reflect the relative abundances in the Sun.  This assumption is also consistent with the overall trend in decades of solar abundance studies: glaring discrepancies between the meteoritic and photospheric abundance scales have historically been resolved in favor of the meteoritic scale.  We will return to this point in Paper II when discussing tests of atmospheres theory.  

The other elements (C, N, O and Ne) cannot be reliably measured in meteorites; neon can only be measured in the extended outer atmosphere of the Sun and in the solar wind. In principle, variations in these elements could be uncorrelated. In order to study the sensitivity of the seismic properties to the abundances, we have generated new opacity tables for each composition and recalibrated the solar model for the new composition. The initial meteoritic, C, N, O and Ne abundances (when applicable) have also been modified for the starting model to include the associated changes in energy generation.  The impact of changes in each of these elements or groups of elements is illustrated in Figure 3 for the AGS05 mixture.  We also include a vector (referred to as photospheric) for the cumulative changes in C, N, O and Ne when varied simultaneously.

Changes in the meteoritic abundances (the dotted arrow in Figure 3) have a strong impact on $Y_{surf}$ but a minimal impact on $R_{CZ}$.  This difference flows naturally from the atomic physics and the absolute abundances.  For high temperatures the more abundant light metals are fully ionized and contribute little to the opacity, while the less abundant heavier metals retain electrons and are still opacity sources.  Changes in the meteoritic abundances therefore affect the core temperature gradient (through the opacity) and the initial helium, which is reflected in the surface helium.  At lower temperatures the more abundant light metals are also strong opacity sources and changes in the meteoritic abundances have little impact on the total opacity at the base of the convection zone.

By contrast, changes in the abundance of the lighter metals as a group (the dot-dashed line in Figure 3) affect both the surface convection zone depth and the surface helium abundance.  When the individual components are examined, there is a clear trend from C to Ne, in the sense that heavier elements have a greater impact on the surface helium.  This is not surprising given the ionization potentials; there is also some effect on the mean molecular weight for the more abundant species such as oxygen.  Changes in carbon and nitrogen clearly have less impact than comparable fractional changes in oxygen, while the most interesting signature of neon is its intermediate response between the CN pattern (convection zone depth only) and the meteoritic pattern (surface helium only).

In order to proceed further, we assume that the CNONe abundances can be varied as a group.  The AGS05 mixture has very similar relative CNO abundances when compared with GS98; C/O and N/O in the AGS05 mixture are only 0.04 and 0.03 dex respectively higher than in GS98, and the Ne/O ratio is only 0.07 dex lower.  This suggests that the atmospheric effects that raise or lower abundances operate on the light metals similarly.  The coronal and solar wind Ne/O ratios are frequently used as abundance indicators (as discussed below), so there is a natural observational scaling between Ne and O.  However, the error in this ratio is larger than the errors in the relative CNO abundances and it is clearly subject to different systematic effects.  We therefore consider two different relative CNONe mixtures (GS98 and AGS05) to determine whether there is a difference in the central photospheric/meteoritic values depending on the mix of light metals.  After this is done, we separately investigate the impact of changes in the relative C, N, and Ne abundances on the overall photospheric and meteoritic abundance scales.

\subsubsection{The Meteoritic-Photospheric Solution}

The clear separation between the group of elements and their differential effects on the solar parameters shown in Figure 2 stimulated us to invert the problem and define the best solution for the photospheric and meteoritic abundances for a given $R_{CZ}$ and $Y_{surf}$. We have calculated the changes induced in these observables for a series of mixtures. Starting with two different initial mixtures (AGS05 and GS98) we have constructed opacity tables and starting models where we held the meteoritic abundances fixed and increased the photospheric abundances by 0.05, 0.1, 0.15, 0.2, 0.25 and 0.3 dex.  We then recalibrated the solar model for each composition. We repeated the operation by changing the meteoritic abundances and holding the photospheric abundances fixed.  Finally, we considered models where the abundances of all elements were held fixed and the C, N, and Ne abundance was varied.  The results are indicated in Figure 4.  Changes in the seismic properties were linear across the range considered for the meteoritic and photospheric changes, while neon abundance variations exhibited a nonlinearity in their response.  In the left panels of the figure we also give a sense of the zero-point shifts in the results from what we view as observationally plausible ($2 \sigma$) changes of 0.14 dex in the neon to oxygen ratio. We then solve 

$$ R-R_{Helio} = \Delta Phot \frac{\delta R_{CZ}}{\delta Phot} + \Delta Met \frac{\delta R_{CZ}}{\delta Met}$$

$$Y-Y_{Helio} = \Delta Phot \frac{\delta Y_{surf}}{\delta Phot} + \Delta Met \frac{\delta Y_{surf}}{\delta Met}$$

for the best estimated abundances changes $\Delta Phot$ and $\Delta Met$ that would reproduce the helioseismic value of $R_{CZ}$ and $Y_{surf}$.  The partial derivatives $dR_{CZ}/dX$ and $dY_{surf}/dX$, where X stands for photospheric or meteoritic, can either be inferred as the slope of the lines in Figure 4 or can be found by interpolation of the partial derivative defined at mid-points, using a 4 points Lagrange scheme.  Both methods yield similar values for the overall meteoritic and photospheric scales.  This procedure gives us the central value. We also derived a range of variation allowed by the errors in the differences between model and theory by mapping the 1 and 2 $\sigma$ error ellipses defined previously in the seismic plane ($\d R_{CZ}- dY_{surf}$) onto the ($\Delta$ Photosperic - $\Delta$ Meteoritic plane).

The best values for the AGS05 mixture are 

$\Delta Phot = 0.207 \pm 0.041~dex$ and

$\Delta Met = 0.056 \pm 0.045~dex$.

In the case of GS98, we obtain 

$\Delta Phot = 0.035 \pm 0.045~dex$ and 

$\Delta Met = 0.0005 \pm 0.037~dex$. 

The uncertainties have been derived from the uncertainties $\sigma_{R_{CZ}}$ and $\sigma_{Y_{surf}}$. This corresponds to 

$[O/H]_{AGS05}=8.87~dex$, $[Fe/H]_{AGS05}=7.51~dex$ and   

$[O/H]_{GS98}=8.86~dex$, $[Fe/H]_{GS98}=7.50~dex$.

We present our abundance solutions in Figure 5. The top panel displays the photspheric and meteoritic solution in the (O-Fe) plane for the relative abundances in the GS98 mixture and the bottom panel gives the results for the relative abundances in the AGS05 initial mixture.

As a cross-check, we have generated 2 opacity tables with these solutions and derived the predicted $R_{CZ}$ and $Y_{surf}$.  Our results are close to the target values.  We obtain $$R_{CZ}^{AGS05+}=0.7138~R_{\odot}~,~ Y_{surf}^{AGS05+}=0.2467$$ and $$R_{CZ}^{GS98+}=0.7133~R_{\odot}~,~Y_{surf}^{GS98+}=0.2482$$

where AGS05+ and GS98+ correspond to the corrected composition.
Given the small differences in the results for AGS05+ compared to the helioseismic results, we estimate that the numerical uncertainty in the derived solution is of order +0.006 dex, or small compared to the error ellipse. 


We find it striking that interiors theory combined with seismic data can provide such a stringent constraint on the composition of the Sun, comparable to the formal error estimates of the most precise claimed absolute spectroscopic measurements.  Although the derived iron/meteoritic abundance scale is consistent with the published central values of both mixtures, the derived oxygen/photospheric abundance scale is clearly incompatible with AGS05 (formally, at the $15~\sigma$ level.)  We therefore conclude that if the AGS05 abundance scale is correct, the stellar interiors models must have an unidentified error source corresponding either to a large underestimate of the uncertainty in the included physics or missing physics.  We can identify no such plausible culprit in the interiors models.

\subsubsection{Effects of Changes in the O-Ne Ratio}

We have repeated the analysis with changes in the C, N, and Ne abundances relative to the other photospheric species.  In the case of C and N we only perform a perturbation analysis, as the changes required for a low oxygen/high CN solution to the seismic problem are unphysical.  However, large changes in neon have been suggested as a possible seismic solution (Antia \& Basu 2005; Bahcall, Serenelli \& Basu 2005a), and there are claims that the solar oxygen might be underestimated using data from solar analogs (Drake \& Testa 2005).  We therefore investigated the impact of significant selective increases in neon.  Reproducing the scalar constraints with the AGS05 CNO abundances held fixed requires a large increase of +0.64 dex in Ne and a slight compensating reduction of -0.013 dex for the meteoritic abundance scale. Our result is very similar to Antia \& Basu (2005), who quoted a change of +0.63 dex based on envelope models.  Although this solution is degenerate seismically with the normal neon abundance mixtures in the previous section, it is not identical when the sound speed profile as a whole is considered.  In Figure 6 we illustrate the sound speed differences between three different solar models consistent with the scalar seismic constraints.  The agreement between the scaled GS98 and AGS05 relative mixtures and the solar sound speed profile are impressively good.
However, the AGS05 CNO/high Ne solution exhibits striking departures from the measured sound speed profile in the radiative core of the Sun.  We therefore conclude that although a high neon abundance can satisfy the convection zone properties of solar models, it does not provide a good fit to the overall sound speed profile of the Sun.  As a result, adopting a high neon abundance alone would not solve the problem of the conflict between the low claimed solar oxygen abundance and stellar interiors theory plus helioseismology.

Furthermore, recent studies of the Sun itself indicate that there are serious problems with adopting a very large increase in the derived neon abundance (Young 2005, P.R., astro-ph/0510264; Schmelz et al. 2005).  We conclude that neon does contribute significantly to the uncertainty in the derived seismic solar oxygen abundance, but that drastic neon abundance variations are neither observationally likely nor seismically favored solutions to the overall problem.

However, more modest changes in carbon, nitrogen, and neon are both observationally plausible and would have some impact on the derived results.  
We therefore modified the abundances of these three elements by 0.1 dex separately and computed the changes in the seismic properties, and thus the difference between data and theory, that would result.  We then re-derived the best fit photospheric and meteoritic abundance scales that would be obtained, and took the differences as a measure of the sensitivity of the overall results to changes in the mix of lighter metals.   
Our final result capturing the correlations between the best abundance indicators are

$O/H = 8.86 \pm 0.041 - ~0.198~d(C)~-~0.135~d(N)~-~0.351~d(Ne)$

$Fe/H = 7.50 \pm 0.045 + ~0.038~d(C)~+~0.014~d(N)~-~0.038~d(Ne)$

Where d(C), d(N), and d(Ne) are the logarithmic differences between the abundances of the species in question relative to oxygen and those found in the AGS05 mixture. Because O and Fe are not perfectly orthogonal variables when used to solve for $R_{CZ}-Y_{Surf}$, they are not completly independant. Therefore the pertubation in the abundance in O and Fe provided by the changes in C,N and Ne abundances generates a residual error of the order of $9\times10^{-4}$ dex in Fe corresponding to an additional error of $10^{-5}$ in $R_{CZ}$ (for a case where +0.1dex is applied to C, N and Ne at the same time). 
The observational errors in the Ne/O, C/O, and N/O abundances are respectively 0.06 dex, 0.05 dex, and 0.06 dex.  If we treat these as uncorrelated, the results above imply an additional uncertainty in the derived Fe and O of 0.003 and 0.025 respectively.  Based on this conservative exercise, we conclude that the total errors in Fe and O are 0.045 and 0.048, which we adopt for our purposes.

\section{Discussion}

The absolute heavy element abundance scale has important implications for astrophysics.  A variety of problems, from the stellar age scale to chemical evolution and tests of the theory of stellar structure and evolution, rely heavily on our knowledge of abundances.  We believe that the recent work on stellar atmospheres theory (as summarized in AGS05) represents a serious effort to improve fundamental stellar observables that requires careful testing by other methods.  Stellar interiors theory provides stringent tests of proposed revisions in other areas of astronomy, and in this paper we demonstrate that theoretical solar models can provide stringent tests of the solar abundances that are consistent with helioseismic data and interiors physics.

We have undertaken a comprehensive error analysis and conclude that the predictions of the interiors models are inconsistent with the low heavy element abundances derived from the latest generation of theoretical model atmospheres at high statistical significance.  Our overall results on seismic agreement are in accord with the Monte Carlo simulation of Bahcall, Serenelli, \& Basu (2005b) that appeared on astro-ph as this manuscript was being completed.  We find a more statistically significant difference than those authors for the surface helium abundance because we include mixing in our models, which is a real physical effect, and have a different method for estimating systematic errors.
The disagreement in convection zone depth is found to be highly significant in both studies.

The sensitivity of the seismic model properties to the data, however, is high enough in our view that this result can be extended to infer the composition of the Sun consistent with seismology when solar models are required to reproduce the surface convection zone properties.  The derived absolute abundances, and their associated errors, have a mild sensitivity to the mix of light metals.  The overall picture that emerges is that the interiors models predict a mixture that is comparable to, or mildly metal-rich relative to, the GS98 abundances.
Large deviations in specific abundances for individual elements, such as neon, have a negative signature in the sound speed profile even if they can reproduce the surface constraints.  We believe this work quantifies the abundance problem and sheds some light on what can and cannot be reliably inferred from helioseismic data.

We are therefore left with a situation analogous to that in the early phases of the solar neutrino problem, namely that we have established that the predictions of atmospheres theory lead to a disagreement with interiors theory that substantially exceeds the currently estimated errors in the interiors theory.  We do not address in this paper the question of what problems, if any, there are in the model atmospheres calculations.  We do believe that it is important to provide a test of the uncertainties in the atmospheres theory in order to determine how significant these differences are, and Paper II is devoted to that question.  For the purposes of this paper, there are two features of the atmospheres problem that we do wish to draw attention to.  The first is that there are real zero-point differences between abundances derived from different model atmospheres codes that are significant at the level of disagreement that we are considering.  It would be valuable to examine the abundances that would be derived from the Asplund et al. (2000) convection treatment for other underlying atmospheres calculations.  The second is that the model atmospheres are also subject to internal consistency tests, in the sense that different techniques and spectral lines can be used to estimate the abundance of the same species.  Trends with excitation potential in the derived abundances from 3D model atmospheres (seen for solar OH r-r lines in Asplund et al. 2004) and discrepancies between solar atomic and molecular indicators for the same species (Allende Prieto, Asplund, \& Fabiani Bendicho 2004) are clearly present in the published 3D model atmospheres, and they may indicate that the errors in atmospheric abundances have been substantially underestimated.

\acknowledgments We dedicate this paper to the memory of John Bahcall.  MP would like to thank Sarbani Basu for discussions during the time this manuscript was being prepared. FD would like to thank Claude Zeippen and Mike Seaton for comments and fruitful discussions.

\clearpage
\input Tab1.tex

\clearpage
\input Tab2.tex
\input Tab3.tex


\begin{figure}
\plotone{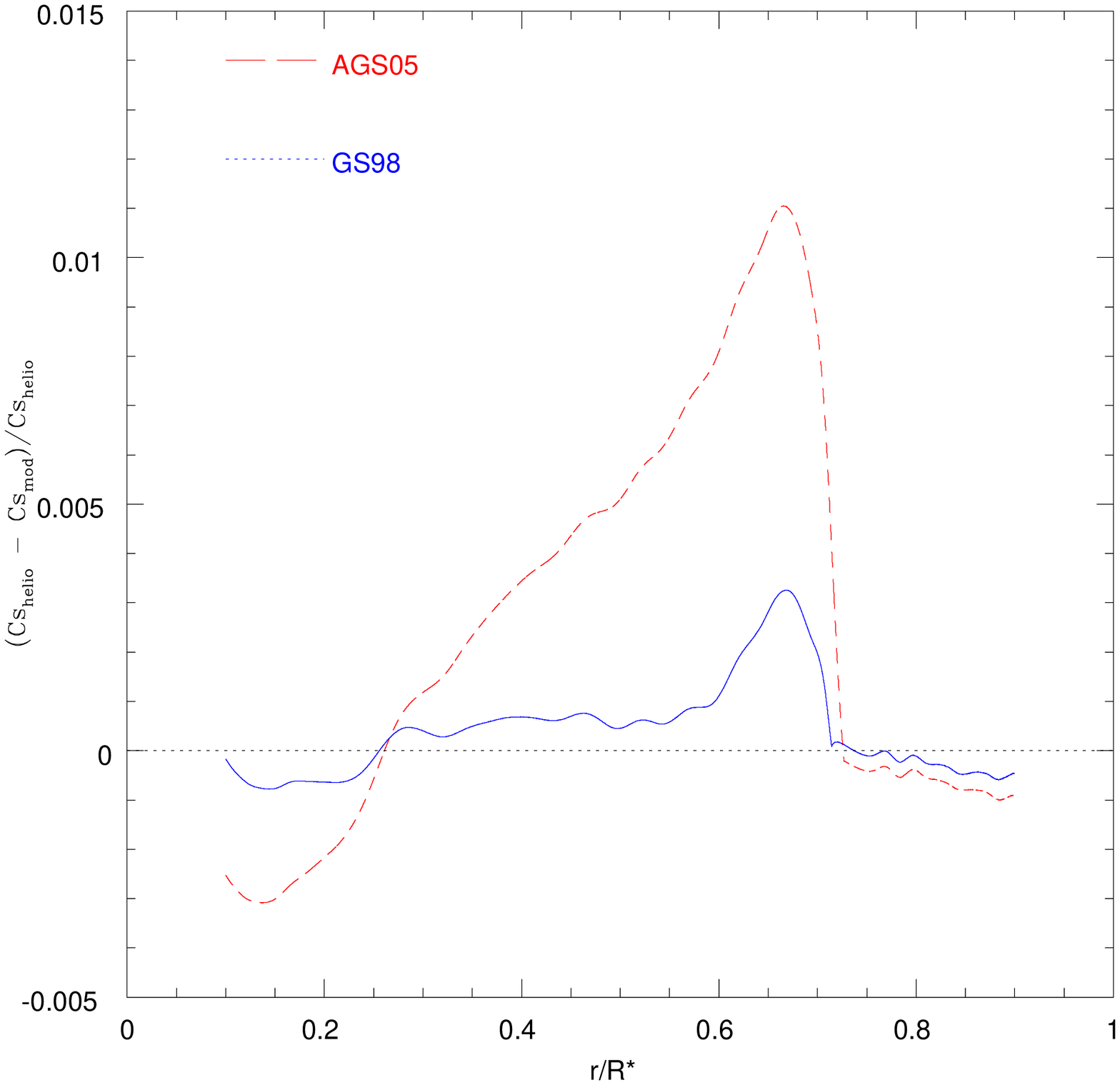}
\caption{Predicted Sound speed profiles for 2 different compositions.}
\end{figure}

\begin{figure}
\plotone{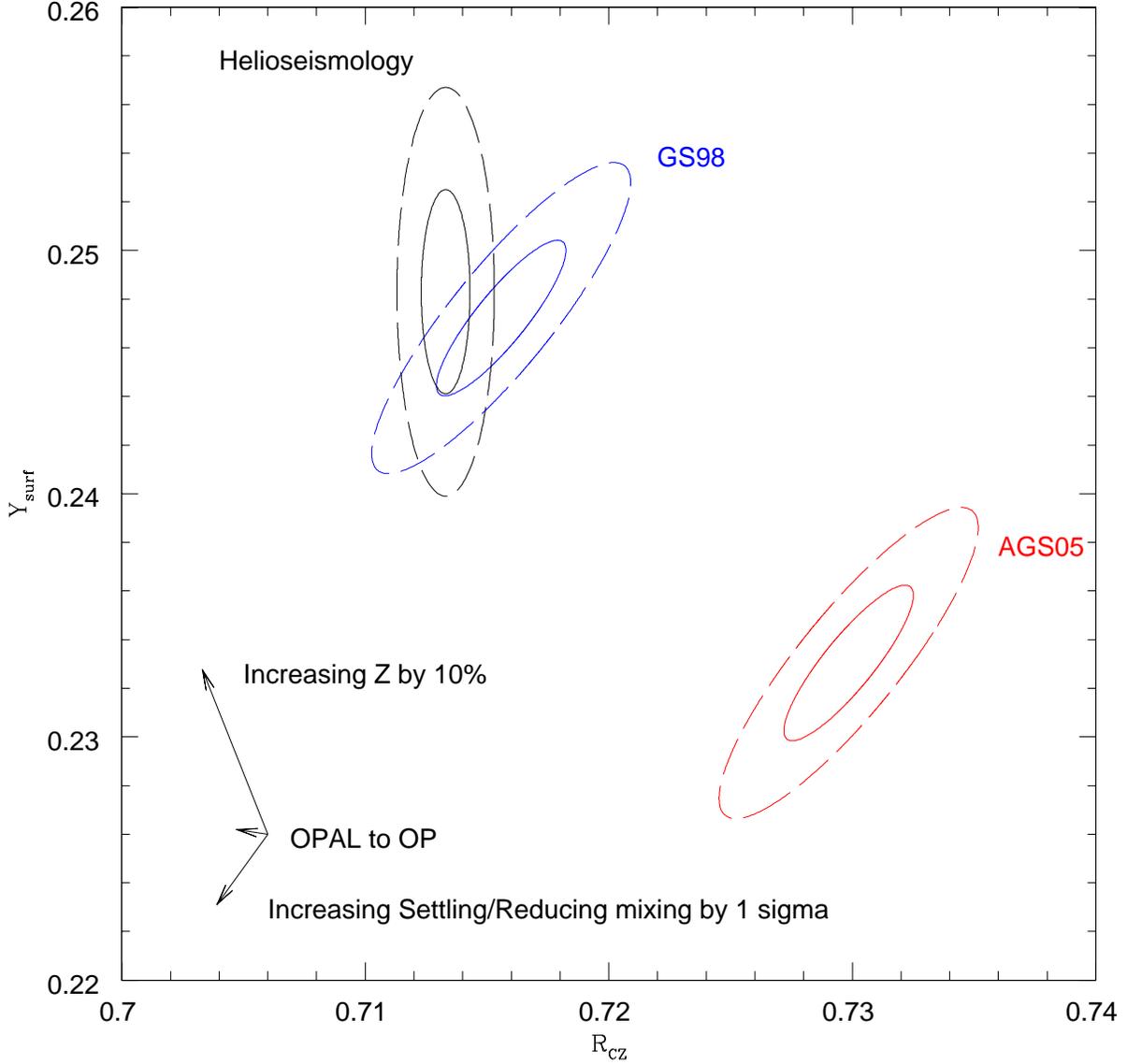}
\caption{ Theoretical and Observational value of $R_{CZ}$ and $Y_{surf}$ with the 
corresponding 1 and 2 $\sigma$ error ellipses. The arrows represent the amplitude
and the direction of the change produced in $R_{CZ}$ and $Y_{surf}$ when the input 
physics is modified. Three cases are presented: An increase of 10\% in the metallicity,
replacing the OP opacity by the OPAL opacity, and changing the level of settling and mixing 
(see text for details). The observational error ellipses are centered the preferred helioseismic values, while the theoretical error ellipses are centered 
on the best seismic values for both the GS98 mixture and the AGS05 mix respectively.}
\end{figure}

\begin{figure}
\plotone{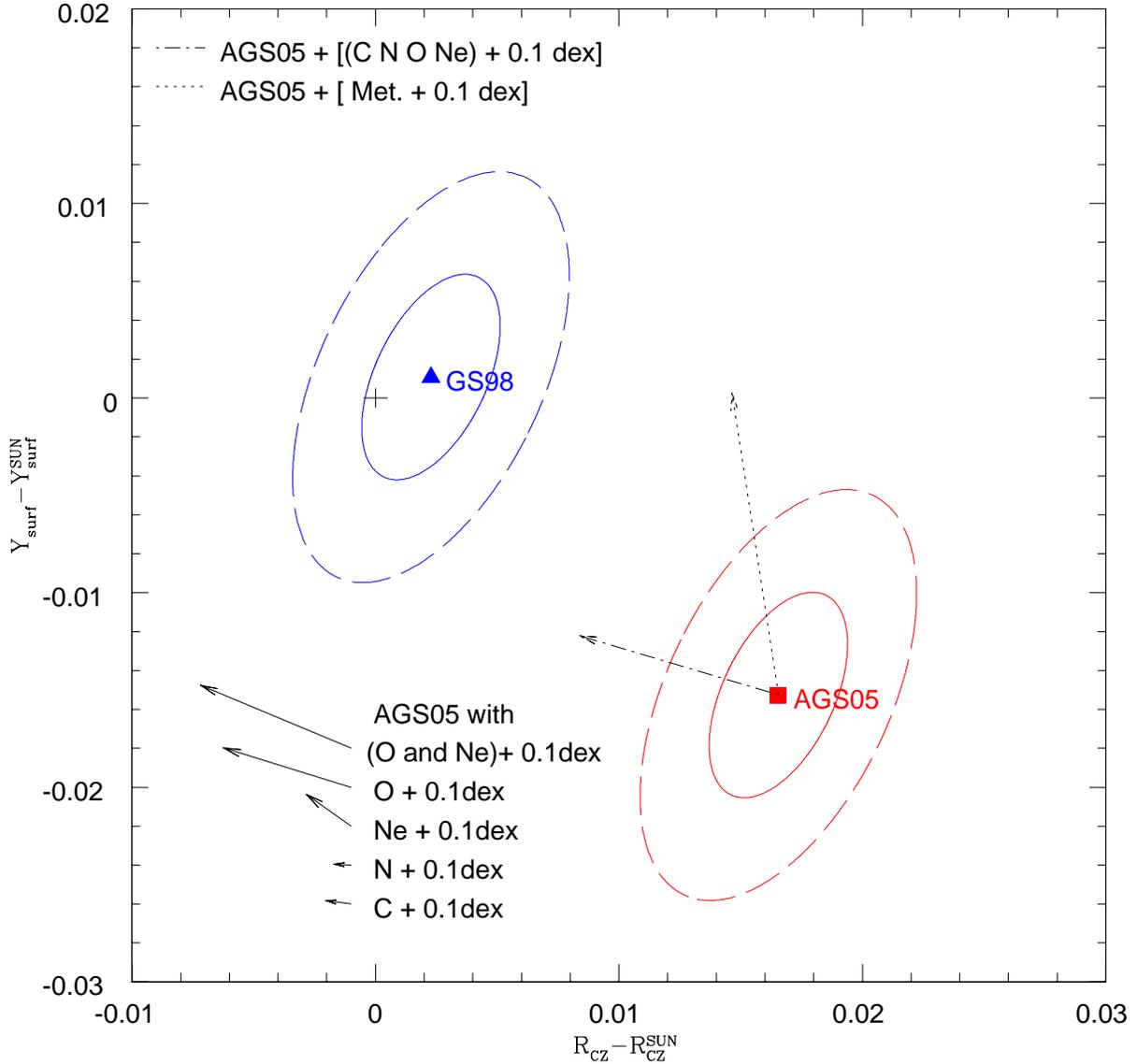}
\caption{ Difference [Theory - Observation] for $R_{CZ}$ and $Y_{surf}$. The ellipses
represent the 1 and 2 $\sigma$ total errors (sum of theoretical and observational which enter 
in the difference). The triangle corresponds to the value predicted using GS98 mixture
and the square those obtained using AGS05 composition. The arrows represent the amplitude
and direction of the change in this difference when the initial composition is modified. 
For the two arrows attached to the AGS05 values, the photospheric (dot-dash) 
and meteoritic (dot) abundances have been increased by 0.1 dex. On the bottom 
left corner, individual elements abundances have been changed, one at a time. 
From top to bottom, (O,Ne) +0.1 dex, O + 0.1 dex, Ne + 0.1 dex, N + 0.1 dex and  
C + 0.1 dex . }
\end{figure}
\begin{figure}
\plotone{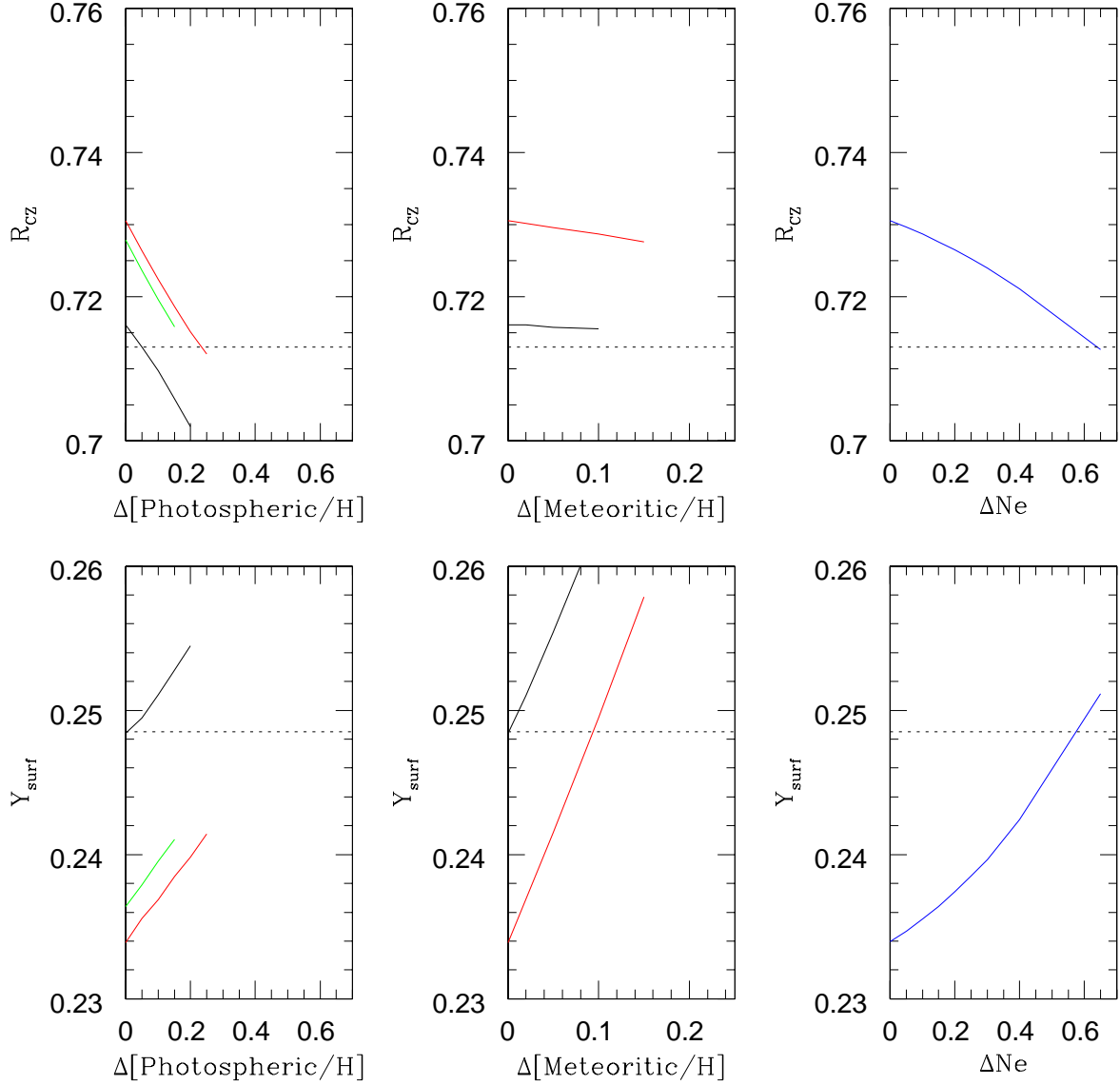}
\caption{ Predicted values of $R_{CZ}$ and $Y_{surf}$ when the composition is modified. The top panels represent the model $R_{CZ}$ as a function of increases in 'photospheric' (C,N,O and Ne) abundances (left), meteoritic abundances (middle) and the neon abundance (right).  The bottom panels are illustrate the sensitivity of $Y_{surf}$ to abundance changes.  Different lines within each panel represent different choices for the starting abundances.  Solid black: changes applied to GS98 mixture, Solid red: changes applied to AGS05 mixture, Green, same as red plus an extra increases of Ne abundances of 0.14dex (dash).}
\end{figure}
\begin{figure}
\plotone{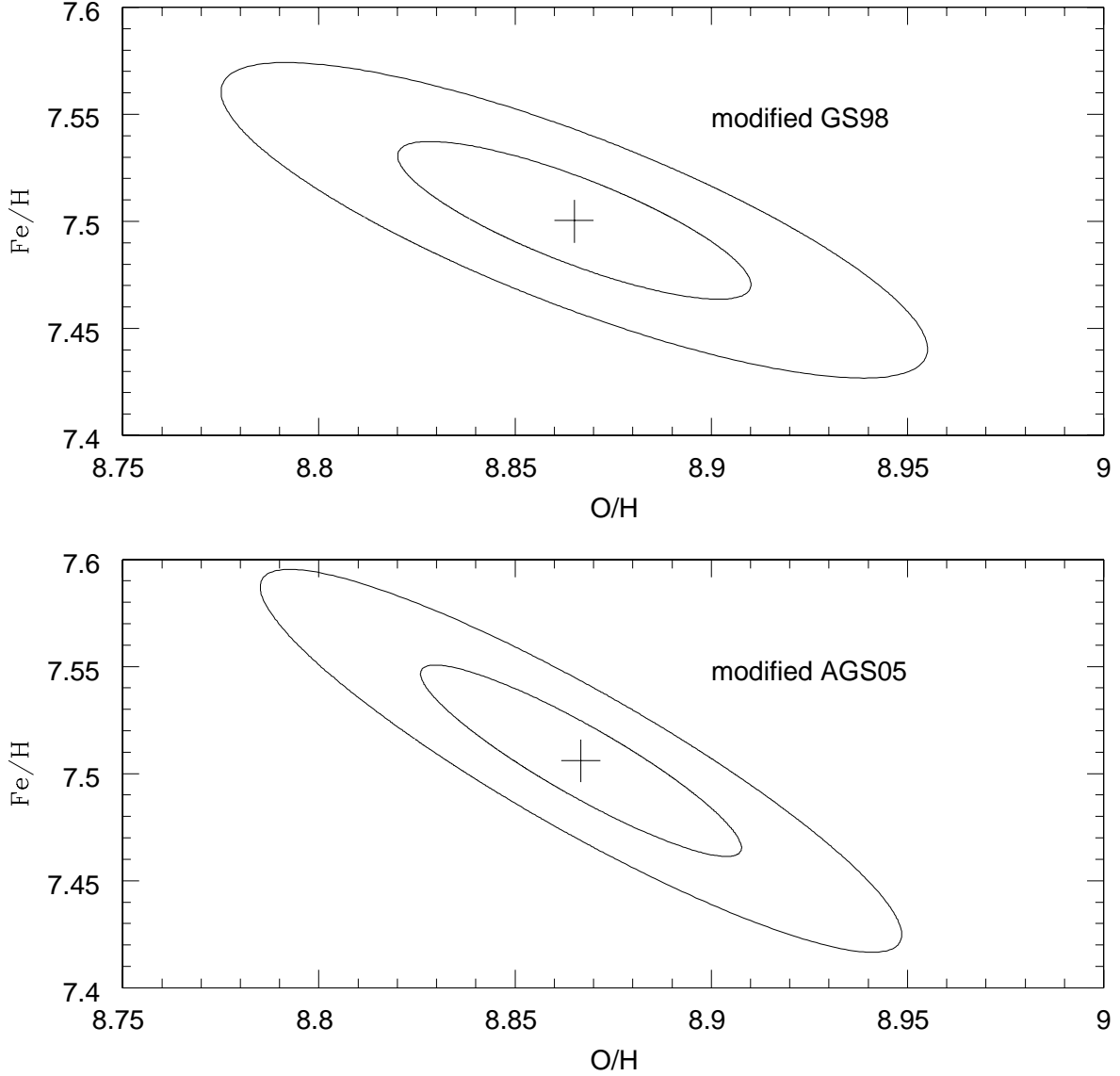}
\caption{ Oxygen and Iron abundances needed to fit $R^{Sun}_{CZ}$ and $Y^{Sun}_{surf}$.
The ellipses represent the domain of composition compatible with 1 and 2 $\sigma$ error in the
difference [prediction - Observation] for $R^{Sun}_{CZ}$ and $Y^{sun}_{surf}$. The top panel pr
esents the best fit when the changes (photospheric as a whole and meteoritic as a whole) are ap
plied to the GS98 composition. The bottom panel presents the best fit when the modifications ar
e applied for the relative light metal abundances in the AGS05 composition. }
\end{figure}

\begin{figure}
\plotone{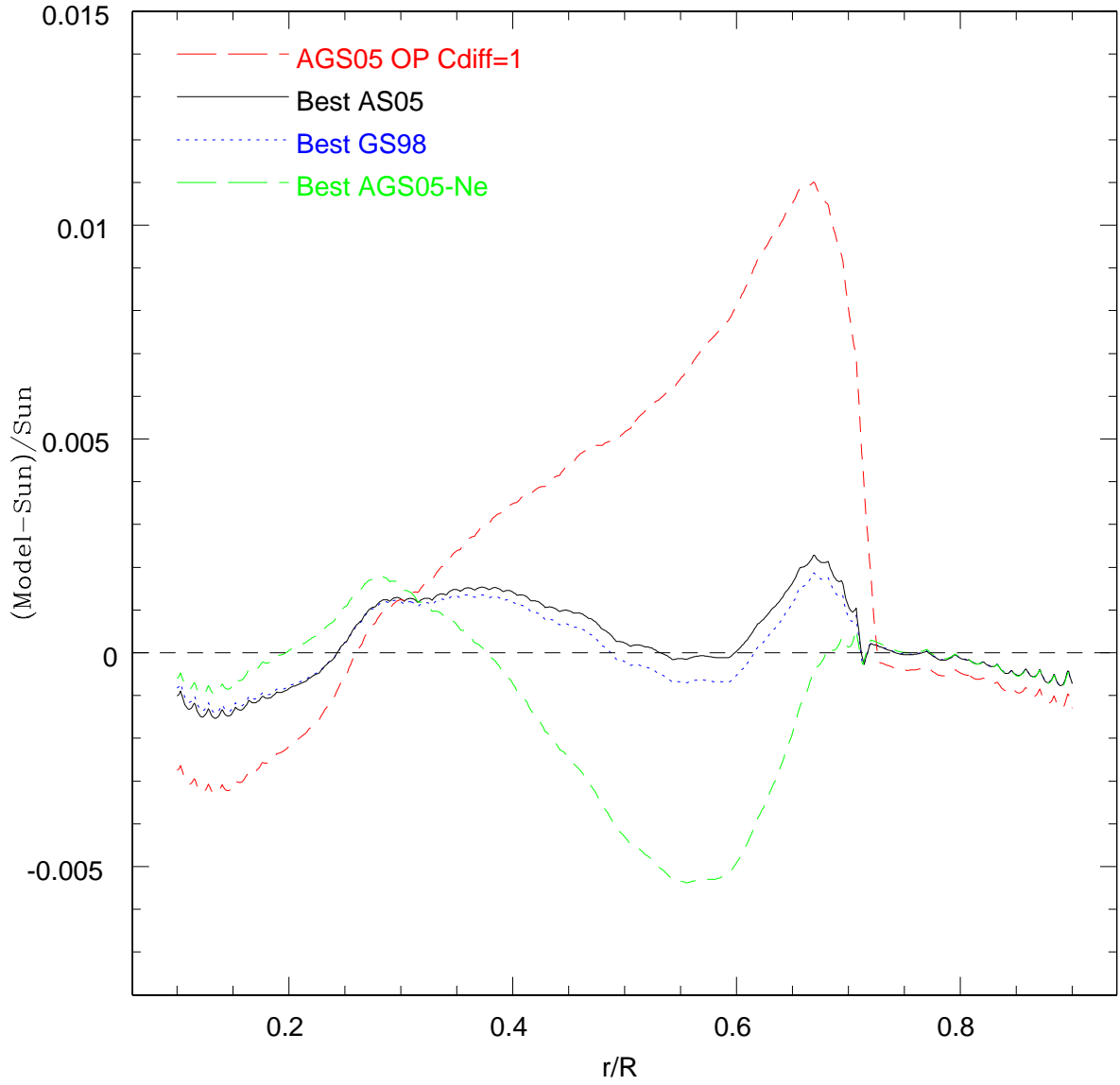}
\caption{Predicted Sound speed profiles for different solar models.} 
\end{figure}

\end{document}

%% file: Tab1.tex
 \begin{table}
\centering
\begin{tabular}{crr}
\hline
\noalign{\smallskip}
 Element & GS98  & AGS05 \\
\noalign{\smallskip}
            \noalign{\smallskip}
\hline
\noalign{\smallskip}
H & 12.000 & 12.000\\
He & 10.930 & 10.930\\
C & 8.520 & 8.390\\
N & 7.920 & 7.780\\
O & 8.830 & 8.660\\
Ne & 8.080 & 7.840\\
Na & 6.330 & 6.270\\
Mg & 7.580 & 7.530\\
Al & 6.490 & 6.430\\
Si & 7.560 & 7.510\\
P & 5.560 & 5.400\\
S & 7.200 & 7.160\\
Cl & 5.280 & 5.230\\
Ar & 6.400 & 6.180\\
K & 5.130 & 5.060\\
Ca & 6.350 & 6.290\\
Ti & 4.940 & 4.890\\
Cr & 5.690 & 5.630\\
Mn & 5.530 & 5.470\\
Fe & 7.500 & 7.450\\
Ni & 6.250 & 6.190\\

\noalign{\smallskip}
\hline\\
\end{tabular}
\\\caption {Initial compositons . }                                              

\end{table}

%% file: Tab2.tex
 \begin{table}
\centering
\begin{tabular}{rcccc}
\hline
\noalign{\smallskip}
Source of error & Central value \& $\sigma$ & $\sigma$&$\Delta R_{CZ}~10^{-3}$ & $\Delta Y_{surf}~10^{-3}$ \\
\noalign{\smallskip}
            \noalign{\smallskip}
\hline
\noalign{\smallskip}
Luminosity  &$3.8418\times10^{33}~ergs\pm0.4\% $ & r & $-0.05/+0.06$ & $+0.39/-0.38$ \\
Age & $4.57\pm0.01~Gyr$          & r & $-0.08/+0.11$      	& $-0.17/+0.12$ \\
$R_{\odot}=695.508~Mm$ & $R_{\odot}^{ref.}=695.98~Mm$  & s & $-0.10$ & $+0.02$   \\
EOS (OPAL 1998)& ref.(OPAL 2001)         & s & $\pm0.12$ & $\pm0.52$  \\
$r_{1,1}$ & +2.1\%/-1.3\% & r & -0.9/+0.64 		& +0.63/-0.42 \\
$r_{3,3}$ & $\pm 7.5\% $  & r & $\pm 0.10$ 		& -0.08/+0.09 \\
$r_{3,4}$ & $\pm 9.4\% $  & r & -0.23/+0.23		& +0.22/-0.24  \\
Low T opacity (Sharp)(a)& ref. (Alexander)    & s & $\pm0.10$ & $\pm0.10$    \\
$\kappa_R$ (OPAL)& ref. (OP)     & s & $\pm0.65 $	& $\pm0.10$ 	\\
Settling+Mixing &+/-16\%  & r & $-2.12/+2.16$    	& $-2.98/+2.93$ \\
diferential Settling &+/-10\%& r &+0.24/-0.21		& +0.59/-0.64\\
Numerics (b)	&	  & r &+1.00/0.00		&$\pm0.00$	  \\
\noalign{\smallskip}
\hline
\end{tabular}
\\\caption{Different sources of error in the determination of $R_{CZ}$ and 
$Y_{surf}$}
in the solar calibration. s = systematic and r = random.
(a) Boothroyd \& Sackmann 2003 (ApJ 583:1004)
(b) Bahcall, Serenelli \& Pinsonneault 2004 (ApJ 614:464)
\end{table}

%% file: Tab3.tex
 \begin{table}
\centering
\begin{tabular}{crr}
\hline
\noalign{\smallskip}
 Initial mixture & $[O/H]$  & $[Fe/H]$ \\
\noalign{\smallskip}
            \noalign{\smallskip}
\hline
\noalign{\smallskip}
AGS05 & 8.864 $\pm$ 0.041  & 7.506 $\pm$ 0.045\\
GS98 & 8.865 $\pm$ 0.045  & 7.501 $\pm$ 0.037\\
\noalign{\smallskip}
\hline\\
\end{tabular}
\\\caption {Best solutions for oxygen and iron starting from two different initiale
compositions. These value have been obtained when all meteoritic on one hand and 
all photospheric element abundances on the other hand
have been modified in group (see text for details). }                                              

\end{table}